# Possible Constraints on Neutron Electric Dipole Moment from Pulsar Radiation

C Sivaram

Indian Institute of Astrophysics

**Abstract:** Even if only a small fraction of neutron dipole moments are aligned in a neutron star, observed pulsar radiation loses provide a stringent limit on the neutron electric dipole moment of $<10^{-29}$ ecm, more stringent than best current experimental limits.

Experimental limits on the electric dipole moment (EDM) of elementary particles like neutrons are of continuing interest as it is well known that existence of such a particle parameter violates both parity (P) and time reversal (T) discrete symmetries [1]. If a neutral particle like the neutron [2] has an EDM, this would be a signature for violation of T-invariance (or CP invariance). Phenomena like the neutral kaon decay are well known examples of CP violation [3], the amount of direct violation $\varepsilon' \sim 10^{-7}$. So in principle, the neutron could have an EDM which is small enough to have been experimentally detected so far. (One can easily see that $(\bar{\sigma}.\bar{E})$ reverses sign under both T and P transformations).

The standard model (SM) [4] calculations yield on estimate of $10^{-31} ecm$. Theories [5, 6] of physics beyond the SM gives rise to values as high as $10^{-26} ecm$, which is close to current experimental upper limits to the neutron electric dipole moment [6].

Here we consider the possibility of constraining the neutron EDM by invoking observed limits on pulsar slowdown by emission of magnetic dipole radiation [7]. As is well known, pulsars are rapidly rotating neutron stars, made almost entirely of neutrons (consisting of around $10^{57}$ neutrons) and having large magnetic fields [8]. Typically millisecond pulsars (period of few ms) radiate $\sim 10^{26} W$ of dipole radiation causing their very precise slow down, i.e. $\dot{T} \sim 10^{-18} ss^{-1}$.

Assume a neutron has an EDM $d_n = (ed)$, e is the charge. Suppose we expect one percent of the neutrons to be aligned, say $N \approx 10^{55}$. Then if $\omega$ be the angular frequency of rotation of the pulsar, the electric dipole radiation can be expressed as:

$$P_{ED} = \frac{2}{3c^3} N^2 d_n^2 \omega^4 \quad \ldots (1)$$

(Following from [9]: $P_{ED} = \frac{2}{3c^3} |\ddot{D}|^2$, $\ddot{D}$ in general is the electric dipole moment, second time derivative squared)

This should be smaller than the observed loss of rotational energy [8], due to spin down, i.e.:

$$P_{ED} < I\omega\dot{\omega} \quad \ldots (2)$$

Where $\omega = \frac{2\pi}{T}$, T is the pulsar period.

Putting in the values $(N \approx 10^{55}, \omega = 10^3 s^{-1}, P_{ED} \approx 10^{33} ergs/s)$, implies from equation (1) and (2), that $d_n^2 < 10^{-58}$, or

$$d_n < 10^{-29} ecm \quad \ldots (3)$$

This is two to three orders lower than current experimental limits. One percent alignment of the dipoles is a reasonable assumption, as there are models invoking alignment of all neutron magnetic moments to generate large magnetic fields in the neutron star interior [8]. If we assume alignment of all the neutron moments, then limit on $d_n$ would be $d_n < 10^{-31} ecm$, close to that calculated for the SM model. The value implied by equation (3) already constraints many theories that go beyond the SM.

It is also of interest to note that recently [10, 11] tight constraints have been placed on a dipole vector component of the gravitational field (as $< 10^{-20}$ of the Newtonian field strength) from the millisecond pulsar slow down. These additional long range vector components have been proposed in recent suggestions [12] for modified gravity theories to provide alternative explanations for dark energy (matter).